\documentstyle[preprint,subeqn,eqsecnum,aps,epsfig,floats]{revtex}
\begin{document}                                                 
\newcommand{\be}{\begin{equation}}
\newcommand{\ee}{\end{equation}}
\newcommand{\ba}{\begin{eqnarray}}
\newcommand{\ea}{\end{eqnarray}}
\newcommand{\bc}{\begin{center}}
\newcommand{\ec}{\end{center}}
\newcommand{\vs}{\vspace*{3mm}}
\newcommand{\dis}{\displaystyle}
\newcommand{\bay}{\begin{array}{rcl}}
\newcommand{\eay}{\end{array}}
\def\RN{Reis\-sner-Nord\-str\"{o}m }
\def\rc{\rho_{\rm crit}}
\def\rl{\rho_\Lambda}
\def\rt{\rho_{\rm tot}}
\def\ie{{\it i.e.\;}}
\def\lp{\ell_{\rm Pl}}
\def\mp{m_{\rm Pl}}
\def\tp{t_{\rm Pl}}
\def\tf{t_{\rm FP}}
\def\om{\Omega_{\rm M}}
\def\oa{\Omega_{\Lambda}}
\def\ot{\Omega_{\rm tot}}
\def\tla{\widetilde{\lambda}_\ast}
\def\tom{\widetilde{\omega}_\ast}
\def\luv{\lambda_\ast^{\rm UV}}
\def\guv{g_\ast^{\rm UV}}
\def\lir{\lambda_\ast^{\rm IR}}
\def\gir{g_\ast^{\rm IR}}
\def\lir{\lambda_\ast^{\rm IR}}
\def\gir{g_\ast^{\rm IR}}
\addtolength{\oddsidemargin}{12mm}
\addtolength{\evensidemargin}{12mm}
\addtolength{\textwidth}{-24mm}
\addtolength{\topmargin}{11mm}
\addtolength{\footskip}{11mm}
\addtolength{\textheight}{-22mm}
\renewcommand{\baselinestretch}{1.5}
\textwidth16cm
\setlength{\oddsidemargin}{0cm}
\setlength{\jot}{0.3cm}                    

\begin{titlepage}
\renewcommand{\thefootnote}{\fnsymbol{footnote}}
\renewcommand{\baselinestretch}{1} 

\begin{flushright}
MZ-TH/02-16\\
\end{flushright}   
\begin{center}

{\Large \sc Cosmological Perturbations 
in Renormalization Group Derived Cosmologies}

\vspace{1cm}
{\large 
A. Bonanno} \\
\vspace{0.5cm}
\noindent
{\it INAF -  Osservatorio Astrofisico di Catania,
Via S.Sofia 78, I-95123 Catania, Italy\\
INFN, Via S. Sofia 64, I-95123 Catania, Italy}

\vspace{1cm}
{\large
M. Reuter}\\
\vspace{0.5cm}
\noindent
{\it Institut f\"ur Physik, Universit\"at Mainz\\
Staudingerweg 7, D-55099 Mainz, Germany}
\end{center}                       

\begin{abstract}
A linear cosmological perturbation theory of an almost homogeneous
and isotropic perfect fluid Universe with dynamically evolving Newton constant $G$
and cosmological constant $\Lambda$ is presented. 
A gauge-invariant formalism is developed by means of the covariant approach,  and
the acoustic propagation equations governing the evolution of 
the comoving fractional spatial gradients of the matter density, $G$, and $\Lambda$
are thus obtained. Explicit solutions are discussed in cosmologies  
where both $G$ and $\Lambda$ vary according to  renormalization
group equations in the vicinity of a fixed point.
\end{abstract}
\end{titlepage}                                                 

\section{Introduction} 
The subject of cosmological perturbations 
continues to attract much attention because it is an essential step 
in understanding any theory of cosmological structure formation. 
The simple idea that the observed structure in the Universe has resulted from
the gravitational amplification of small primordial fluctuations works
remarkably well and it must be discussed in any theory of relativistic cosmology.

In inflationary cosmology the presently observed structures in the Universe
are generated by quantum fluctuations during an early de Sitter phase.
The subsequent evolution is classical and depends on the interplay between 
pressure forces, the rate of growth of the expansion factor 
and on the content and the nature of the, yet unknown, dark matter. 

Very recently, an alternative scenario has been 
proposed. In \cite{br1} we discussed a cosmology of the Planck Era, valid
immediately after the initial singularity,  in which 
the Newton constant and the cosmological constant are dynamically 
coupled to the geometry by ``improving'' the Einstein equations with the
renormalization group (RG) equation for Quantum Einstein Gravity 
\cite{mr}.
It has then been shown that in this new scenario a solution to the horizon and 
flatness problem of standard cosmology is possible without the introduction of 
an {\it ad hoc} scalar field. There is also a natural mechanism generating 
primordial density fluctuations with spectral index $n=1$, at least at 
short wavelengths. 

The recent evidence for a non-trivial ultraviolet fixed point 
in Quantum Einstein Gravity \cite{ol,ol2}  has opened the possibility of a non-perturbative
renormalizability of the theory along the lines of the {\em asymptotic safety} 
conjectured by Weinberg \cite{wein}. This has triggered a number
of investigations in black hole physics \cite{bh2,bh1} and cosmology [1]. 
According to the findings of 
\cite{mr,souma,ol,ol2,frank,per,max}
Newton's constant is an asymptotically
free coupling at very high energy scales.   
Given the RG flow near this non-trivial fixed point
it is possible to ``RG improve'' the Einstein field equations by replacing Newton's constant
$G$ and the cosmological constant $\Lambda$ by their scale-dependent (``running'')
counterparts.
The {RG improved} Einstein equations
lead to a mathematically tractable system of evolution equations for the
scale factor, the pressure, the density, $G$, and $\Lambda$. Near the fixed point
explicit solutions are available \cite{br1}. 

A similar mechanism could in principle be operating also far from the Big Bang, in 
a completely different era of the Universe. 
In fact, in the late Universe, the possibility that another, this time infrared
attractive fixed point, might be present in the RG flow for gravity 
and governs its long distance behavior
has been discussed in \cite{br2}. In this case a solution
of the ``cosmic coincidence problem'' \cite{coscon2} arises naturally  
without the introduction of a {\rm quintessence} field.
It can be shown that in the fixed point regime the
vacuum energy density $\rho_\Lambda\equiv\Lambda/8\pi G$
is automatically adjusted so as to equal the matter energy density, 
{\it i.e.} $\oa=\om =1/2$, and that the deceleration parameter approaches $q = -1/4$.
Moreover, an analysis of the high-redshift SNe Ia data leads to the conclusion that 
this {\em infrared fixed point cosmology} is in good  agreement with
the observations \cite{ebe}, and that it is a promising candidate for describing the 
dynamics of the Universe on very large scales.
It is important to note that the experimental determination of $\om$ and $\oa$ is model
dependent. The numbers which are usually quoted, $\om\approx 0.3$ and $\oa \approx 0.7$,
are obtained when one fits the data to the cosmological standard model. Our prediction
$\om=\oa=0.5$ is not inconsistent with this result since we use a different model; in fact,
we find that it fits the supernova data as accurately as the standard model. 

Cosmologies with a time dependent $G$ have been 
discussed in \cite{barrow}, while
homogeneous and isotropic cosmological models 
with variable $G$ and $\Lambda$ have been discussed in \cite{ber} and 
\cite{sys1,cw}. Also for this reason a formalism where 
the evolution of small fluctuations of the density, of $G$, and of $\Lambda$
can be consistently discussed is needed.
The aim of this paper is to present such a formalism 
and to apply it to the case of the RG derived cosmologies.  

We shall follow the so-called covariant approach, pioneered by Hawking \cite{haw},
further extended by Ellis and Bruni \cite{eb}, Jackson \cite{jackson}
and Zimdahl \cite{zim}, in order to provide a gauge-invariant formulation of the problem. 
The gauge issue is in fact a major problem in discussing the evolution 
of the long wavelength 
fluctuations, where the presence of spurious gauge modes can 
be more difficult to avoid \cite{bardeen}. 
The second order acoustic propagation equations, governing 
the comoving fractional spatial gradients of the density 
and of $G$ and $\Lambda$ are then obtained for a class of background spacetimes
with a time-dependent Newton constant and cosmological constant where the matter field
is described by a perfect fluid with conserved energy momentum tensor.    
The properties of the solutions are then discussed
for a class of RG derived cosmologies. It turns out that 
it is possible to have both growing and decaying modes for the 
fractional quantities describing the perturbations of the density, $G$ and 
$\Lambda$, depending on the equation of state. 

In particular, if the late time evolution of the background Universe is governed by 
the infrared fixed point cosmology we shall find that all the relevant perturbations
decay with a power law of the cosmic time or stay constant, {\it i.e.} the 
infrared fixed point cosmology is stable under small deviations 
from perfect homogeneity and isotropy. According to this result, the ultimate fate of the
Universe at very large scales is a sort of ``eternal dilution'' where a 
homogeneous and isotropic state is reached  for $t\rightarrow \infty$.
\section{Gauge-invariant perturbation theory}
In this section we generalize the formalism of refs.\cite{eb,jackson,zim} by allowing 
$G\equiv G(x^\mu)$ and $\Lambda\equiv\Lambda(x^\mu)$ to be scalar functions on
spacetime. 

A ``fundamental observer'' describing the cosmological fluid
flow lines has 4-velocity 
\be\label{2.1}
u^\mu= dx^\mu/ d\tau , \;\;\;\;\;\;\; u^\mu u_\mu = -1
\ee
where $\tau$ is the proper time along the fluid flow lines.
The projection tensor onto the tangent 3-space orthogonal to $u^\mu$ is 
\be\label{2.2}
h_{\mu\nu} = g_{\mu\nu}+u_\mu u_\nu
\ee
with  ${h^\mu}_\nu {h^{\nu}}_\sigma = 
{h^\mu}_\sigma$ and ${h^{\mu}}_\nu u^\nu =0$. 
The covariant derivative of $u^{\mu}$ is
\be\label{2.3}
u_{\mu;\nu}=\omega_{\mu\nu}+\sigma_{\mu\nu}+
{1\over 3}\Theta h_{\mu\nu}-\dot{u}_{\mu} u_{\nu} 
\ee
where  $\omega_{\mu\nu}={h^{\alpha}}_\mu {h^{\beta}}_\nu u_{[\alpha;\beta]}$
is the vorticity tensor, $\sigma_{\mu\nu} 
= {h^{\alpha}}_\mu {h^{\beta}}_\nu u_{(\alpha;\beta)}-{1\over 3}\Theta h_{\mu\nu}$
is the shear tensor, $\Theta = {u^\mu}_{;\mu}$ is the expansion scalar and
$\dot{u}^\mu = {u^{\mu}}_{;\nu} u^\nu$ is the acceleration four-vector; square and round 
brackets denote anti-symmetrization and symmetrization, respectively. 
The Riemann tensor is defined by
$u_{\mu;\lambda\tau}-u_{\mu;\tau\lambda}={R^{\sigma}}_{\mu\lambda\tau}u_{\sigma}$ and
the Ricci tensor by $R_{\mu\nu} = {R^{\tau}}_{\mu\tau\nu}$. With these conventions, 
Einstein' s equations are
\be\label{2.4}  
R_{\mu\nu}-{1\over 2} R g_{\mu\nu}= -\Lambda g_{\mu\nu}+8\pi G T_{\mu\nu}
\ee
where $\Lambda=\Lambda(x^\mu)$ is the position dependent cosmological constant and 
$G=G(x^\mu)$ the position dependent Newton constant.
The energy-momentum tensor $T_{\mu\nu}$ is assumed to be conserved. 
For a perfect fluid it has the 
form
\be\label{2.5}
T^{\mu\nu} = \rho \; u^\mu u^\nu + p\;h^{\mu\nu}
\ee 
The conservation law ${T^{\mu\nu}}_{;\nu}=0$ leads to mass-energy conservation
\be\label{2.6}
{\dot{\rho}\over (\rho + p)} +\Theta=0
\ee
and the equation of motion
\be\label{2.7}
\dot{u}^\mu + {h^{\mu\nu}p_{;\nu}\over \rho + p} = 0.
\ee
The Bianchi identities require the RHS of (\ref{2.4}) to be covariantly 
conserved. This consistency condition together with   
the conservation laws (\ref{2.6}) and (\ref{2.7})
provides the equations for $\Lambda$ and $G$, 
\ba\label{2.8}
&&\dot{\Lambda} + 8 \pi\dot{G}\rho = 0\\[2mm]
&&h^{\mu\nu}\Lambda_{;\nu} - 8\pi p \;h^{\mu\nu} {G}_{;\nu} =0\label{2.8b}
\ea
by projecting along $u^\mu$ and onto the hyperplane orthogonal to $u^\mu$.
The Raychaudhuri equation is obtained with the help of the Einstein 
field equations and of Eq.(\ref{2.3}),
\be\label{2.9}
\dot{\Theta}+{1\over 3}\Theta^2+2(\sigma^2-\omega^2)-
\dot{u}^\mu_{\;\; ;\mu}+4 \pi G (\rho+3p)-\Lambda=0
\ee
where $2\sigma^2 \equiv  \sigma_{\mu\nu}\sigma^{\mu\nu}$ and 
$2\omega^2 \equiv \omega_{\mu\nu}\omega^{\mu\nu}$. The term $ \dot{u}^\mu_{\;\; ;\mu}$
can be rewritten as \cite{zim}
\be\label{2.9b}
\dot{u}^\mu_{\;\; ;\mu}=
-h^{\lambda\tau}\Big ({ {h^\nu}_\lambda p_{;\nu}\over \rho+p} \Big )_{;\tau}+
h^{\lambda\tau}{p_{;\lambda}\over \rho+p}{p_{;\tau}\over \rho +p}.
\ee 
The scalar ${\cal K}$ is defined as
\be\label{2.10}
{\cal K}\equiv 2\sigma^2-{2\over 3}\Theta^2+16\pi G\rho +2\Lambda
\ee
It is possible to show that for zero vorticity,  
$\omega_{\mu\nu}=0$, it coincides with the Ricci scalar $^{(3)}R$ 
of the 3-dimensional hyperplane everywhere orthogonal to $u^\mu$. 
An auxiliary length scale $S(t)$ is introduced as
the solution of the equation
\be\label{2.12}
{\dot{S}\over S}={1\over 3}\Theta.
\ee
Suitable quantities useful to characterize the spatial inhomogeneities
of density, pressure and expansion are, respectively,
\be\label{den}
D_\mu \equiv {S \; {h_\mu}^\nu \;\rho_{;\nu} \over \rho +p},\;\;\;\; 
P_\mu \equiv {S \; {h_\mu}^\nu \; p_{;\nu} \over \rho +p},\;\;\;\; 
t_\mu \equiv {S \; {h_\mu}^\nu \; \Theta_{;\nu} }
\ee
In order to characterize spatial inhomogeneities of $G$ and $\Lambda$ it is
convenient to introduce the following dimensionless quantities:
\be\label{den2}
\Gamma_\mu \equiv \frac{1}{G}\; {{S \; {h_\mu}^\nu \; G_{;\nu}}},\;\;\;\;\;\;\;\;\; 
\Delta_\mu \equiv \frac{1}{\Lambda}\; {S \; {h_\mu}^\nu \; \Lambda_{;\nu}}.
\ee
We then have from (\ref{2.6}) and (\ref{2.7}) 
\ba\label{2.13}
&&{h^\alpha}_\gamma (S {h^\beta}_\alpha \rho_{;\beta}\dot{)}=
{S\over 3} \Theta {h^{\alpha}}_\gamma \rho_{;\alpha}+S\Theta {h^{\alpha}}_\gamma p_{;\alpha}
+S{h^\alpha}_\gamma (\rho_{;\alpha} \dot{)}\nonumber\\ 
&&=-S\Theta  {h^{\alpha}}_\gamma 
\rho_{;\alpha}-S\Theta {h^{\alpha}}_\gamma p_{;\alpha} (p+\rho)
-S {h^{\alpha}}_\beta \rho_{;\alpha}({\omega^\beta}_\gamma+{\sigma^\beta}_\gamma)
\ea
In the first line we used the identity 
$(\rho_{;\alpha}\dot{)}=\dot{\rho}_{;\alpha}
-\rho_{;\beta}{u^\beta}_{;\alpha}$ 
and in the second line (\ref{2.3}). 
Eq.(\ref{2.13}) can now be rewritten as 
\be\label{2.16}
{h^\nu}_\mu\dot{D}_\nu +{\dot{p}\over \rho +p}D_{\mu}+ 
({\omega^\nu}_\mu +{\sigma^\nu}_\mu)D_\nu+t_\mu = 0.
\ee
We can obtain the equation of motion for $t_\mu$ by following similar manipulations: 
\ba\label{2.17}
&&{h^\nu}_\mu\dot{t}_\nu = \dot{S}{h^\nu}_\mu\Theta_\nu  
+S{h^\nu}_\mu\dot{u}_\nu\dot{\Theta}+
S{h^\nu}_\mu (\Theta_{;\nu}\dot{)}\nonumber\\ 
&&=-\dot{\Theta}P_\mu-({\omega^\nu}_\mu +{\sigma^\nu}_\mu)t_\nu-{2\over 3}\Theta t_\mu
-S{h^\nu}_\mu (2\sigma^2-2\omega^2)_{;\nu}+
S{h^\nu}_\mu ({\dot{u}^\tau}_{\;\; ;\tau})_{;\nu}\nonumber\\
&&-4 \pi G(\rho + p) [D_\mu +3P_\mu]
-4\pi G\Gamma_\mu(\rho +p)
\ea
In the last line we have used the Raychaudhuri 
equation (\ref{2.9}) and Eq.(\ref{2.8b}).
We stress that Eq.(\ref{2.16}) and Eq.(\ref{2.17}) are completely
general in the sense that no assumption on the functional form of $G$
or $\Lambda$ has been made. In particular, 
Eq.(\ref{2.17}) coincides with Eq.(27) of \cite{zim} 
for constant $G$ and $\Lambda$. 

Given an equation of state $p=p(\rho)$ 
it is possible to express $P_\mu$ in terms of 
$D_\mu$, but (\ref{2.16}) and  (\ref{2.17}) still do not form a closed system since an
evolution equation for $\Gamma_\mu$ is needed. It can be obtained in the following 
way: from the definition of $\Gamma_\mu$ we have
\ba\label{pina1}
&&{h^\nu}_\mu ({G \Gamma}_\nu \dot{)} =-P_\mu \dot G +S {h_\mu}^\nu 
\dot G_{,\nu} -G \Gamma_{\nu}({\omega^\nu}_\mu +{\sigma^\nu}_\mu)\\[2mm]\label{pina2}
&&{h^\nu}_\mu ({\Lambda \Delta}_\nu \dot{)} =-P_\mu \dot \Lambda +S {h_\mu}^\nu 
\dot \Lambda_{,\nu} -\Lambda \Delta_{\nu}({\omega^\nu}_\mu +{\sigma^\nu}_\mu)
\ea
where we have used Eq.(\ref{2.3}). From Eq.(\ref{2.8}) and Eq.(\ref{2.8b}) we
get instead 
\ba\label{pina3}
&&{h^\nu}_\mu ({\Lambda \Delta}_\nu \dot{)}=8\pi \dot p \;G \Gamma_\mu +
8\pi p \;{h^\nu}_\mu ({G \Gamma}_\nu \dot{)}\\[2mm]
&&h^{\mu\nu}\dot \Lambda_{,\nu}+8\pi h^{\mu\nu}\rho_{,\nu}+
8\pi \rho \;h^{\mu\nu}\dot{G}_{,\nu}=0\label{pina4}
\ea
Therefore by using (\ref{pina2}) in (\ref{pina3}) and using (\ref{pina4}) in the result,
(\ref{pina1}) becomes
\be\label{pina5}
{h^\nu}_\mu ({G \Gamma}_\nu \dot{)} =-{\dot p \over \rho +p} G \Gamma_\mu
-\dot G D_\mu - {G\over \rho +p}\Gamma_\nu  
({\omega^\nu}_\mu +{\sigma^\nu}_\mu)
\ee
This is the equation for $\Gamma_\mu$ we were looking for.

From now on we shall assume the background Universe 
to be homogeneous and isotropic, {\it i.e.}
$\omega_{\mu\nu}=\sigma_{\mu\nu}=\dot{u}_\mu=0$. In this Universe we consider 
small perturbations
of the motion of the fluid and consequently, up to first order
in the inhomogeneities, the factors multiplying the quantities $D_\mu$, $P_\mu$,  
$t_\mu$ and $\Gamma_\mu$ 
in (\ref{2.16}) and (\ref{2.17}) refer to the background. Thus the linearized
equation for $D_\mu$ becomes
\ba\label{2.18}
&&{h^\nu}_\mu \dot{D}_\nu +{\dot{p}\over \rho +p}D_{\mu}+t_\mu = 0
\ea
and the linearized equation for $t_\mu$ reads
\ba\label{2.19}
&&{h^\nu}_\mu\dot{t}_\nu=- 
\dot{\Theta}P_\mu-{2\over 3}\Theta t_\mu+
S{h^\nu}_\mu ({\dot{u}^\tau}_{\;\; ;\tau})_{;\nu}
-4\pi G (\rho + p) [D_\mu +3 P_\mu]-4\pi G\Gamma_\mu (\rho +p)
\ea

These equations can be further simplified if we specify the 
factors in front of the quantities $D_\mu$, $P_\mu$, for the background spacetime. 
In particular, in a homogeneous Universe with variable $G$ and $\Lambda$
the relevant equations for the background evolution read
\begin{subequations}
\ba\label{2b1}
&&{\cal K}=2(-{1\over 3}\Theta^2+8\pi G \rho +\Lambda)= {^{(3)}}R\\
&&\dot{\Theta}+12\pi G (\rho +p) = {{^{(3)}}R }/2\label{2b2}\\
&&\dot{\Lambda} + 8 \pi\dot{G} \rho=0\label{2b3}
\ea
\end{subequations}
These equations imply that we may set $S(t)=a(t)$. In fact, we
observe that Eq.(\ref{2b1}) and Eq.(\ref{2b2}) 
are the $00$-component and $ii$-component of 
Einstein's equation, respectively,
provided we identify $S(t)$ with the
scale factor $a(t)$ of the Robertson-Walker line element
\be\label{met}
ds^2=-dt^2+a(t)^2\Big [{dr^2\over 1-Kr^2}+r^2(d\theta^2+\sin^2\theta \;d\phi^2)\Big ]
\ee
For $K=0$ the $3$-spaces of constant cosmological time $t$ are flat, 
and for $K=+1$ and $-1$ they are spheres and pseudo-spheres, respectively.
The 3-curvature of the $t=const$ hyperplane is 
\be\label{2.21b}
{^{(3)}}R={6K\over a^2}.
\ee
As a consequence of the above results, Eq.(\ref{2.19}) now reads
\be\label{2.22}
{h^\nu}_\mu\dot{t}_\nu
=-{{^{(3)}}R\over 2}P_\mu -{2}{\dot a\over a}\;t_\mu
-4\pi G (\rho + p)D_\mu +S{h^\nu}_\mu ({\dot{u}^\tau}_{\;\; ;\tau})_{;\nu}
-4\pi G\Gamma_\mu(\rho +p)
\ee
Eq.(\ref{2.9b}), up to linear order, implies  \cite{zim}: 
\be\label{2.23}
S{h^{\tau}}_\nu (\dot{u}^\mu_{\;\; ;\mu})_{;\tau} = -{\nabla^2 \over a^2}P_\nu
\ee
where $\nabla^2$ is the Laplacian on the maximally symmetric 3-space. 
For an equation of state $p=p(\rho)$ we may write  
\be\label{2.24}
\dot{p} = - c^2_s\;\Theta\; (\rho +p), \;\;\;\;\;\; P_\mu =c^2_s \;D_\mu.
\ee
where $c_s=(\partial p/\partial \rho)^{1/2}$ denotes 
the velocity of sound in the medium. 
Therefore Eq.(\ref{2.22}) becomes
\be\label{2.22b}
{h^\nu}_\mu\dot{t}_\nu
=-{2}{\dot a\over a}\;t_\mu
-\Big [ 4 \pi G (\rho + p) +c^2_s \;\frac{3K}{a^2} +
c^2_s {\nabla^2\over a^2}\Big ]D_\mu 
-4 \pi G\Gamma_\mu(\rho  + p)
\ee
By combining (\ref{2.22b}) with (\ref{2.18}) and (\ref{2.24})
and noticing that ${h^\mu}_\tau({h^\nu}_\mu \dot{D}_\nu\dot{)}
={h^\nu}_\tau\ddot{D}_\nu-\dot{u}_\tau \dot{u}^\nu {D}_\nu
={h^\nu}_\tau\ddot{D}_\nu$ up to second order, 
we get the desired differential equation for the fractional density
which is of second order in $t$:
\ba\label{2.25}
&&{h_\mu}^{\nu} \ddot{D}_\nu+({2}-3\;c^2_s){\dot a\over a}\;
{h_\mu}^{\nu} \dot{D}_\nu\nonumber \\
&&-\Big [ 3({c^2_s} \dot{)}{\dot a\over a} +
4 \pi G [(\rho-3 p)+2\Lambda]c^2_s+
4 \pi G(\rho+p) + {c^2_s} \;{\nabla^2 \over a^2} \Big ]
D_\mu\nonumber\\
&&-4 \pi G \Gamma_\mu(\rho +p)=0
\ea
The corresponding first order evolution equation 
for $\Gamma_\mu$ reads according to  Eq.(\ref{pina5})
\be\label{gonda}
{h_\mu}^\nu (G \Gamma_\nu\dot{)} =
c^2_s \;\Theta\;G \Gamma_\mu-
\dot{G} D_\mu 
\ee
Once the solution for $\Gamma_\mu$ is obtained also 
$\Delta_\mu $ is known. Eq. (\ref{2.8b}) yields the simple relationship
\be\label{londa}
\Delta_\mu = {8\pi \;p \;G \over \Lambda}\; \Gamma_\mu
\ee

The coupled system (\ref{2.25}), (\ref{gonda}), (\ref{londa}) 
are the general gauge-invariant equations which govern the first order 
perturbations of cosmologies with variable $G$ and $\Lambda$.
They are one of our main results. As an immediate consequence of (\ref{londa})
we observe that $\Delta_\mu=0$ in any Universe with $p=0$ so that 
$\Lambda$ is always spatially constant in this case.  
\section{RG derived cosmologies}
The idea of using the RG flow equation in gravity is borrowed 
from particle physics where the RG improvement is a standard device in order to add, 
for instance, the dominant quantum corrections 
to the Born approximation of a scattering cross section. However, instead of improving 
{\it solutions}, in \cite{br1,br2} the more powerful improvement of the basic 
{\it equations} has been discussed. We shall now briefly review 
the main results for an homogeneous and isotropic Universe.

We assume that there is a fundamental scale dependence of Newton's constant which is governed by an 
exact RG equation for a Wilsonian effective action whose precise nature needs not to be specified
here. At a typical length scale $\ell$ or mass scale $k = \ell^{-1}$ those ``constants''
assume the values $G(k)$ and $\Lambda(k)$, respectively. In trying to ``RG-improve'' Einstein's 
equation the crucial step is the identification of the scale $\ell$ or $k$ which is relevant in 
the situation under consideration \cite{berbell}. In cosmology, the postulate of homogeneity and
isotropy implies that $k$ can depend on the cosmological time only so that the scale dependence is
turned into a time dependence:
\be
G(t)\equiv G(k=k(t)),\;\;\; \Lambda(t) \equiv \Lambda(k=k(t))\label{3.d}
\ee
In principle the time dependence of $k$ can be either explicit or implicit via the scale factor:
$k=k(t,a(t),\dot{a}(t),\ddot{a}(t),\cdots )$. In ref. \cite{br1} we gave detailed arguments as to why
the purely explicit time dependence 
\be\label{ide}
k(t) = {\xi} /{t}
\ee
with $\xi$ being a positive constant of order unity is the correct identification. In a nutshell,
the argument is that, when the age of the Universe is $t$, no (quantum) fluctuation with a frequency
smaller than $1/t$ can have played any role yet. Hence the integrating-out of modes
(``coarse graining'') which underlies the Wilson renormalization group should be stopped 
at $k\approx 1/t$.
Moreover, other plausible cutoffs, such as $k\approx H(t)$, are equivalent to (\ref{ide}) under very general
conditions. (See \cite{br1} and \cite{br2} for further details.)

Once the 
RG trajectory 
 $k\mapsto (G(k), \Lambda(k))$ is known, the system (2.26) together with 
(\ref{3.d},\ref{ide}) can be solved. 
In the following we focus on the vicinity of a fixed point where simple 
analytical solutions can be found \cite{br1,br2}.

To be precise, we assume that within a large basin of attraction the 
dimensionless quantities $g(k)\equiv k^2 G(k)$ 
and $\lambda(k) \equiv \Lambda(k)/k^2$ 
get attracted either towards an ultraviolet (UV) fixed point for $k\rightarrow \infty$ or towards 
an infrared (IR) fixed point for $k\rightarrow 0$. While the physics of these two situations
is quite different (early vs. late Universe) they are rather similar mathematically 
and we can discuss the two cases in parallel 
\footnote{If both of the fixed points are present one could have a very 
symmetric scenario with respect to the ``birth'' and ``death'' of the Universe
where the cosmological evolution amounts to a cross-over
from the UV to the IR fixed point. See ref.\cite{liouv} for a 
similar cross-over in 2D gravity.}. 

As for the actual derivation of those fixed points
from the RG equations, a non-gaussian UV 
fixed point is known to exist in the Einstein-Hilbert
truncation of pure Quantum Einstein Gravity \cite{mr,souma,ol,frank,per}
and in more general truncations \cite{ol2}. Recently it was  also found
within the 2 Killing-vector reduction of Quantum Einstein Gravity \cite{max}. 
For the time being,  
the existence of an IR fixed point  cannot be deduced yet from a (truncated)
RG equation because from a technical point
of view it is very difficult to follow the RG flow very far towards the IR.
The reason is that a reliable description of 
IR physics most probably requires truncations
containing nonlocal invariants. 
In the second paper
of \cite{frank} a first progress was made in this direction. Using a simple,
mathematically tractable nonlocal truncation it was shown that there is a
scale invariant IR fixed point at which the size of the universe is
completely unrelated to the value of the bare cosmological constant. The problem is that 
this truncation has an interpretation only within {\it euclidean} quantum gravity
and does not direcly imply the IR fixed point postulated in the present paper. 
Nevertheless, this investigation suggests that the late Universe, at very large scales,
may be regarded as a kind of scale free ``critical phenomenon''. A natural way of
implementing this general physical picture within Lorentzian gravity is by means of a 
fixed point for $g$ and $\lambda$.

It is also conceivable that the postulated IR fixed point is of a completely classical
nature \cite{classav}  or that the quantum effects of matter fields play an 
important role \cite{sola}. 

In the vicinity of a fixed point $(g_\ast,\lambda_\ast)$
the evolution of the dimensionful $G$ and 
$\Lambda$ is approximately given by
\be\label{8}
G(k) = \frac{g_\ast}{k^2},\;\;\;\; \Lambda(k) = \lambda_\ast \; k^2
\ee
From
(\ref{8}) with (\ref{ide}) we obtain the time dependent Newton constant and cosmological
constant: 
\be\label{9}
G(t) = g_\ast \xi^{-2} \; t^{2},\;\;\;\; \Lambda(t) = \frac{\lambda_\ast\xi^2}{t^2}
\ee
The power laws (\ref{9}) are valid for $t\searrow 0$ (UV case)
or for $t\rightarrow \infty$ (IR case), respectively. 
If we use these functions $G(t)$ and 
$\Lambda(t)$ in the coupled system (2.26), 
its solution gives us the scale factor 
$a(t)$ and the density $\rho (t)$ of the ``RG improved cosmology''. 

In the case of a spatially flat Universe $(K=0)$ and the equation 
of state $p=w\rho$, which we shall consider in 
this paper, the system (2.26) 
with (\ref{9}) has the following 
one-parameter family of solutions:
\begin{subequations}
\ba\label{11a}
&&a(t) = \Big [ \Big ({3\over 8}\Big )^2 (1+w)^4 \; 
g_\ast \lambda_\ast \; {\cal M} \Big ]^{1/(3+3w)}\; t^{4/(3+3w)}\\[2mm]
&&\rho(t) = {8\over 9\pi (1+w)^4 \; g_\ast \lambda_\ast}\; {1\over t^4}\label{11b}\\[2mm]
&&G(t) = {3\over 8}\;(1+w)^2 \; g_\ast \lambda_\ast \; t^2\label{11c}\\[2mm]
&&\Lambda(t) = {8\over 3(1+w)^2}\; {1\over t^2}\label{11d}
\ea
\end{subequations}       
Note in particular that $a\propto t$ for $w=1/3$ 
and $a\propto t^{4/3}$ for $w=0$.                                                     
Apart from the constant $w$ and the product 
$g_\ast \lambda_\ast$, the solution (3.5) depends
only on a single constant of integration, ${\cal M}$, whose value affects only the
overall scale of $a(t)$. 
Numerically it equals $8\pi \rho(t) [a(t)]^{3+3w}\equiv {\cal M}$
which, like in standard cosmology, is a conserved quantity.
Introducing the critical density 
\be\label{12}
\rc (t) \equiv \frac{3}{8\pi G(t)}\;\Big ( \frac{\dot{a}}{a} \Big )^2
\ee
we find for any value of $w$, $g_\ast\lambda_\ast$, and ${\cal M}$ that 
$\rc (t) = 2 \rho (t)$ and $\rho_\Lambda (t) = \rho (t)$. Hence 
\be\label{13}
\rho=\rho_\Lambda
=\frac{1}{2}\rc
\ee
Thus the total energy density $\rt \equiv \rho +\rl$ equals precisely the critical one:
$\rt (t) = \rc (t)$. 

The exact equality, at any time,  of the matter energy 
density $\rho$ and the vacuum energy density 
$\rl$ is a nontrivial prediction of the fixed point solution. In terms of the
relative densities, 
\be\label{15}
\om = \oa = \frac{1}{2}, \;\;\;\;\;\;\;\;\; \ot =1
\ee
Also the Hubble parameter of the solution (3.5)
\be\label{16}
H \equiv \frac{\dot{a}}{a} = \frac{4}{3+3w}\; \frac{1}{t}
\ee
and its deceleration parameter
\be\label{17}
q \equiv -\frac{a\,\ddot{a}}{\dot{a}^2} = \frac{3w-1}{4}
\ee
are independent of $g_\ast$, $\lambda_\ast$ and ${\cal M}$ \cite{berto}. (It can be shown
that the standard formula for $q$ in terms of the relative densities
continues to be correct for {\it all} solutions of the improved system (2.6) 
with an arbitrary RG trajectory:
$q = \frac{1}{2} \; (3w+1)\;\om-\oa$.)
\section{Stability of fixed point cosmologies}
In Section II we derived a general system of equations governing the 
dynamics of small perturbations about a cosmology with variable $G$ and 
$\Lambda$. In the following we apply this formalism to the fixed point 
background cosmology of (3.5).
We write 
\be\label{2.26}
D^\mu = \delta^n (t) \; \Psi^\mu_n \;\;\;\;\;\; \Gamma^\mu = \gamma^n (t) \;\Psi^\mu_n
\ee
where $\Psi^\mu_n$ is an eigenfunction of the Laplacian 
operator $\nabla^2$ with (negative) eigenvalues $\nu_n^2$:
\be\label{har}
-\nabla^2\;\Psi^\mu_n = \nu^2_n\;\Psi^\mu_n
\ee
For an equation of state of the type $p=w\rho$ we then obtain
the following closed system for the perturbed quantities:
\ba\label{2.27}
&&\ddot\delta^n+({2}-3w){\dot a\over a} \dot\delta^n-
\Big [4\pi G \rho (1-w)(3w+1) +2w\Lambda \Big ]\delta^n 
+w \;\nu^2_n\;\delta^n=4\pi\rho G\gamma^n(1+w)\\[2mm]
&&\dot\gamma^n =  3 \;w\;{\dot{a}\over a}\gamma^n  - {\dot  G \over G}\gamma^n
-{\dot  G \over G}\delta^n .\label{zonno}
\ea

We consider this system in the long wavelength limit for which $\nu_n\approx 0$.
Its solutions are simple power-laws.
If we set 
$\delta = A t^\alpha$ and $\gamma = B t^\beta$
then, by direct substitution in (\ref{2.27}), one finds that $\beta = \alpha$, and 
$\alpha$ is obtained by solving a cubic equation. It has three real roots:
\be \alpha_1 = \frac{4 w}{1+w} \;\;\;\;\;\; \alpha_2 = \frac{2(3w-1)}{3(w+1)}
\;\;\;\;\;\; \alpha_3 = \frac{w-3}{w+1}\label{4.6}
\ee     

In the late Universe governed by an IR fixed point, $p=0$ {\it i.e.} $w=0$
should be a good approximation to the equation of state. In this case the exponents  
are $\alpha = (0, -2/3, -3)$. It is not necessary to invoke the long wavelength limit
in order to arrive at this result; for $w=0$ the eigenvalues $\nu^2_n$ drops out
from  (\ref{zonno}) so that we are left with the same equation for perturbations at any 
wavelength. As a consequence, there are no growing modes of the $\rho-$ and $G-$perturbations.
Furthermore, as we mentioned already, $\Lambda$ is strictly spatially constant in a 
$p=0$ Universe so that there are no inhomogeneous $\Lambda-$perturbations at all.
We conclude that in an IR-fixed point Universe there exist no linear perturbations
which would drive the evolution away from the ideal homogeneous and isotropic 
state. This implies that there is no Jeans-type instability giving rise to the 
formation of structures {\it on cosmological scales}. 

This last restriction stems from the fact that the above derivation made essential
use of the cutoff identification (\ref{ide}) which assumes that the relevant momentum
scale $k$ is given by the inverse cosmological time. For a perfectly homogeneous 
and isotropic Universe this is essentially the unique choice  \cite{br1,br2}.
Therefore the above analysis should apply to perturbations with proper wavelengths 
on the Megaparsec scale or beyond where the Universe starts looking homogeneous  
and isotropic. The dynamics of perturbations on smaller length scales is much harder
to analyze. The reasons are: {\it (i)} It is not clear how to identify the running scale 
$k$ in terms of physical quantities since in presence of inhomogeneities or
anisotropies $1/t$ is presumably not the only relevant scale. {\it (ii)} Smaller lengths
correspond to higher values of $k$ for which the RG trajectory might still be far away from
the IR fixed point. Thus more detailed knowledge about the trajectory $k\mapsto (G(k),\Lambda(k))$
is necessary. 

In order to get a first understanding of what could happen at small length scales 
let us consider the following ``toy-model'' RG trajectory. Let us assume that 
for $k$ below a certain critical value, $G(k)$ and $\Lambda(k)$ run according to 
the fixed point law (\ref{9}) and they are approximately constant for 
$k$ well above this critical value. This behavior is motivated by the fact that 
from laboratory to galactic scales, say, we do not see any variation of $G$ and
$\Lambda$. In this model the evolution of the homogeneous and isotropic 
{\it background} is still described by the fixed point cosmology: $k$ is small, 
the RG-trajectory is close to the fixed point, and $k\propto 1/t$ is the 
unique cutoff identification. However, if ``lumps'' form due to the gravitational
attraction and if they are sufficiently small then their size can correspond to 
a $k-$value smaller than the critical one so that $G$ and $\Lambda$ are constant 
across these structures. As a consequence, a density perturbation is {\it not} 
accompanied by a perturbation (spatial inhomogeneity) of $G$ in this case. 

Small scale perturbations of this type are described by Eq.(\ref{2.27}) with 
$\gamma^n\equiv 0$ on the RHS,  
\be\label{ztn}
\ddot\delta^n+( {2}-3w){\dot a\over a}\; \dot\delta^n-
\Big [{4 \pi G \rho} (1-w)(3w+1) +2w\Lambda \Big ]\delta^n 
+w \; \nu^2_n \; \delta^n=0
\ee
Formally this equation is the same as in standard cosmology, but now
the background evolution is given by the fixed point solution (3.5). 
If we set $\delta_n(t) =A t^\alpha$, we obtain
\be \alpha_1 = \frac{5(3w-1)-\sqrt{81 w^2+138 w +73}}{6(1+w)}
\;\;\;\;\;\;\;
\alpha_2 = \frac{5(3w-1)+\sqrt{81 w^2+138 w +73}}{6(1+w)}\label{4.8}
\ee
and there is always a growing mode and a decaying mode, as in the standard case.
In deriving (\ref{4.8}) we used the long wavelength limit $\nu^2_n=0$ 
in (\ref{ztn}). As we are interested in small structures this is not
necessarily always allowed, but for the most relevant case $w=0$  
the eigenvalue $\nu^2_n$ drops out again so that (\ref{4.8}) is 
indeed valid for small scale perturbations. For $w=0$ the density 
perturbations grow approximately as
\be\label{4.9}
\delta(t) \propto t^{0.59} \propto a^{0.44}
\ee

In \cite{br1} we developed a cosmology of the Planck era immediately after the 
Big Bang which was governed by an UV fixed point. In this scenario 
$w=1/3$ is a distinguished choice for the equation of state. From Eq.(\ref{4.6}) 
we read off that in this ``radiation dominated Planck era'' there exist
perturbations which grow as $\delta(t)\propto t \propto a(t)$. 
Interestingly, for $w=1/3$, Eq.(\ref{ztn}) with $\nu^2_n=0$ yields a
qualitatively similar result; from (\ref{4.8}) we obtain growing modes  
with $\delta(t) \propto t^{\sqrt{2}} \propto a^{\sqrt{2}}$.

\section{conclusion}
We have presented a general covariant framework to investigate the evolution
of cosmological perturbations in Robertson-Walker spacetimes 
with variable $G$ and $\Lambda$. 
In comparison to earlier work
\cite{barrow,ber,sys1,cw,alpha} on cosmologies with a time dependent $G$, $\Lambda$
and possibly fine structure constant $\alpha$ the new feature of our model is that
the time dependence of $G$ and $\Lambda$ is a secondary effect which derives from a
more fundamental scale dependence. In a typical Brans-Dicke type theory, say, the dynamics
of the Brans-Dicke field $\Phi =1/G$ is governed by a standard local lagrangian with 
a kinetic term $\propto (D_\mu \Phi)^2$. In our approach there is no
simple lagrangian description of the $G$-dynamics. It rather arises from an RG
equation for $G(k)$ and a cutoff identification $k=k(x)$. From the point of view of the
gratitational field equations, $G(x)$ has the status of an external scalar field.

Our main interest was to analyze a special type of spacetimes where the evolution
of $G$ and $\Lambda$ in the background Universe is governed by a RG flow
near a fixed point. In this case the 
flow is very simple and it is possible to obtain
explicit solutions in a model independent way. In fact, as for the
IR fixed point governing the late-time behavior of the Universe, no further assumptions
beyond the very fixed point hypothesis need to be made. In particular, 
the IR fixed point is not necessarily due to quantum gravity effects as is the
UV fixed point, it could have a purely classical origin 
(for instance within a classical averaging scenario \cite{classav}).  

We performed a detailed stability analysis of the fixed point solution (3.5)
which was first discussed in \cite{br1,br2}. With the equation of state 
$p=0$, appropriate for the late Universe, we found that the IR fixed point
cosmology is stable against the formation of large scale inhomogeneities.
As a consequence, structure formation on cosmological scales stops by the time
when the Universe  enters the fixed point regime. We also found that perturbations
on small length scales still can grow even in the fixed point epoch. 
As a very rough estimate, we expect the dividing line between large- and small-scale
perturbations to be of the order of the Hubble length. 

It also turned out 
that if the evolution of the Universe immediately after the initial singularity at 
$t=0$ is described by a UV fixed point solution (3.5) with $w>0$, there is always an
amplification of the small disturbances generated during the quantum gravity era.
In particular for $w=1/3$ density perturbations grow as $\delta_n \propto a$. 
These fluctuations would then emerge as ``primordial'' density perturbations
at the end of the Planck era and follow the subsequent evolution 
according to the standard Friedmann-Robertson-Walker 
dynamics. 

It is clear, though, that the classical isotropy problem of the Big Bang
cosmology is not solved by the IR fixed point postulate. A satisfactory solution to 
this problem would involve starting from a generic inhomogeneous and anisotropic Universe
and showing that it evolves into an almost homogeneous and isotropic one long before
the time of decoupling. It is intriguing to speculate that RG effects play a crucial
role here. In this context it is an encouraging first result that the 
UV fixed point cosmology is free from the particle horizon present in the radiation
dominated standard model \cite{br1}.

Clearly more work is needed in order to confront the (IR) fixed point cosmology with 
the observations. As a first step we compare in \cite{ebe} our predictions to the recent
high-redshift supernova data. 
It turns out that, at least as far as these data are concerned,
the fixed point model is phenomenologically viable. In ref.\cite{ebe} we also extend the
fixed point model proper by assuming that the fixed point epoch is preceded by an era with 
a constant $G$ and $\Lambda$. In this model, because of the accelerated $t^{4/3}$-expansion
after the transition, perturbations first evolve inside the horizon, then they become larger
than the horizon. If we assume that the
dividing line between ``large'' and ``small'' perurbations is indeed of the order $H^{-1}$,
then, in the fixed point regime, perturbations first grow inside the horizon with the
exponents (\ref{4.9}) and at some point they reach the asymptotic constant regime with (\ref{4.6}).
This behavior is similar to the one found in $\Lambda$-dominated cosmologies 
and it is a consequence of the cosmic-no hair theorem \cite{tava}.
We shall come back to the extended model elsewhere.

We also hope
to implement this formalism in a more realistic framework, taking into account  the 
presence of dark matter in a multifluid description, in a $N$-body simulation where 
the background solution is given by the IR fixed point cosmology.
Also, it will be important to
analyze the model in the context of the recent data on the
cosmic microwave background radiation and the cluster
density data.

\section*{Acknowledgements}
We would like to  thank V.Lukash, 
G.Esposito, P.Forg\'acs, C.Kiefer, C.Rubano, M.Niedermaier and M.Sereno 
for very interesting discussions. A.B. also acknowledges the warm 
hospitality of the Physics Departments of the University of Napoli and of 
Mainz University, where part of this work was written. 
M.R. would like to thank the Astrophysical Observatory of Catania
for the hospitality extended to him.

\end{document}